\documentclass[a4paper, 12pt]{article} 

\usepackage{amsfonts} 
\usepackage{amsmath}

\begin{document} 

\title{Finite-amplitude inhomogeneous plane waves \\of   
exponential type  \\in incompressible elastic materials.} 

\author{Michel Destrade}
\date{1999}
\maketitle

\bigskip

\begin{abstract} 
It is proved that elliptically-polarized finite-amplitude
inhomogeneous plane waves may not propagate in an isotropic elastic
material subject to the constraint of incompressibility. 
The waves considered are harmonic in time and exponentially 
attenuated in a direction distinct from the direction of 
propagation. 
The result holds whether the material is stress-free or 
homogeneously deformed.
\end{abstract} 

\newpage

\section{Introduction} 
 
The problem of wave propagation in a medium is addressed 
mathematically by seeking solutions for the displacement of a 
particle to a `wave equation' characteristic of the medium.
Among possible solutions, harmonic forms for the displacement 
are of great interest because any linear combination of harmonic 
waves is also a solution of the wave equation.
Harmonic waves vary sinusoidally with time and distance, as 
they travel with constant speed and unchanged profile in a 
fixed direction.
They are called `homogeneous plane waves' because the displacement
field is homogeneous in the planes orthogonal to the direction of
propagation.

However, for certain physical situations, such as gravity waves,
surface waves, or reflection and refraction of waves, an 
attenuation of the amplitude occurs in a direction distinct 
from the direction of propagation.
Thus arises the need to find `inhomogeneous plane waves' 
solutions to the wave equation.

A simple form for the displacement is that of a vector field 
$\textbf{u}(\textbf{x}, t)$ which varies sinusoidally
with frequency $\omega$ in the direction of a vector 
$\textbf{S}^+$ and is attenuated exponentially in the direction 
of another vector $\textbf{S}^-$, so that 
$\textbf{u}(\textbf{x}, t)$ 
is the real part of the complex quantity 
$ e^{- \omega  \textbf{S}^- \textbf{.x}}  
\{ \textbf{A} e^{i \omega (\textbf{S}^+ \textbf{.x} - t)} \}$, 
where \textbf{A} is the amplitude of the wave. 
The complex vector (or `bivector' \cite{BoHa93}) 
$\textbf{S} = \textbf{S}^+ +i \textbf{S}^-$ is called the 
`slowness bivector' and its real and imaginary parts describe 
the `planes of constant phase' 
($\textbf{S}^+ \textbf{.x}=\mathrm{constant})$ 
and the `planes of constant amplitude' 
($ \textbf{S}^- \textbf{.x}=\mathrm{constant}$). 
When $ \textbf{S}^+ $ and $\textbf{S}^-$ are  parallel, 
the wave is homogeneous; otherwise, it is inhomogeneous.
Similarly, \textbf{A} is a bivector, whose real and
imaginary parts either are parallel (linear polarization) or have
distinct directions (elliptical polarization).

In this note, we place ourselves in the context of finite
elasticity. 
Specifically, we are interested in the propagation of plane
waves of exponential type in incompressible elastic materials. 
It has been shown that finite-amplitude homogeneous plane waves 
(with linear or elliptical polarization) may propagate in 
deformed incompressible materials 
(Green \cite{Gree63}, Currie \& Hayes \cite{CuHa69}, 
Boulanger \& Hayes \cite{BoHa92}). 
Also, small-amplitude elliptically-polarized inhomogeneous
plane waves propagating in a deformed incompressible material 
have received much attention 
(e.g. Hayes \& Rivlin \cite{HaRi61}, Flavin \cite{Flav63},
Belward \cite{Belw73},  Belward \& Wright \cite{BeWr87}, 
Borejko \cite{Bore87}, Boulanger \& Hayes \cite{BoHa96}, \ldots). 

Here we show that \textit{elliptically}-polarized 
\textit{inhomogeneous} plane waves of \textit{finite} 
amplitude may \textit{not}  propagate in any
incompressible material, whether deformed or not.

\section{Proof}
 
For a deformation bringing a material point from position 
\textbf{X} in the reference configuration to position 
\textbf{x} in the current configuration, the deformation gradient 
\textbf{F} is defined by 
\begin{equation}  
\textbf{F}= \frac{\partial \textbf{x}}{\partial\textbf{X}}.  
\end{equation}

Because of the incompressibility constraint, any deformation of the
material must be isochoric, so that, at all times, we have
\begin{equation} \label{det}
\mathrm{det} \, \textbf{F} = 1.
\end{equation}

Consider the propagation of an elliptically-polarized 
inhomogeneous plane wave of finite amplitude, which we choose 
to be of an exponential form. 
Thus, if \textbf{A} is the amplitude bivector of the wave
and $\textbf{S} = \textbf{S}^+ +i \textbf{S}^-$ is the  
slowness bivector with associated frequency $\omega$, then the 
wave is given by 
\begin{equation} \label{x} 
\textbf{x}  = \textbf{X} 
+  \left\{ 
\textbf{A}e^{i\omega(\textbf{S.X}-t)} +  
\widetilde{\textbf{A}}
e^{-i\omega(\tilde{\textbf{S}}.\textbf{X}-t)}
\right\}, 
\end{equation}
where \textbf{A} is of finite magnitude and the tilde denotes the
complex conjugate.

The deformation gradient \textbf{F} associated with this 
deformation is given by 
\begin{equation}  
\textbf{F}=\frac{\partial \textbf{x}}{\partial\textbf{X}} 
= \textbf{1} +  \omega \left\{ i \textbf{A} \otimes \textbf{S} 
e^{i \omega ( \textbf{S}.\textbf{X}-t)} -
i \widetilde{\textbf{A}} \otimes \widetilde{\textbf{S}}
e^{-i\omega(\tilde{\textbf{S}}.\textbf{X}-t)}\right\}.  
\end{equation}

Then $J = \mathrm{det} \, \textbf{F}$ is given by
\begin{multline} \label{J}
J=  \, 1 
+  \omega \left\{ i(\textbf{A}.\textbf{S}) 
 e^{i\omega(\textbf{S}.\textbf{X}-t)} 
-i (\widetilde{\textbf{A}}.\widetilde{\textbf{S}}) 
e^{-i\omega(\tilde{\textbf{S}}.\textbf{X}-t)}\right\}  
 \\   
- \omega^2 \left\{ (\textbf{A}. \widetilde{\textbf{S}}) 
(\widetilde{\textbf{A}}. \textbf{S}) - (\textbf{A}. \textbf{S}) 
(\widetilde{\textbf{A}}. \widetilde{\textbf{S}}) \right\} 
 e^{i\omega(\textbf{S}-\tilde{\textbf{S}}).\textbf{X}}. 
\end{multline} 

However, because $J=1$ at all times by \eqref{det}, we must have 
$\textbf{A}.\textbf{S} = 0, \, 
\widetilde{\textbf{A}}. \widetilde{\textbf{S}} =0, \,
(\widetilde{\textbf{A}}. \textbf{S})
(\textbf{A}.\widetilde{\textbf{S}})  
- (\textbf{A}. \textbf{S})
(\widetilde{\textbf{A}}.\widetilde{\textbf{S}}) = 0$, 
which implies 
\begin{equation} \label{A.S}
\left\{ 
\begin{array}{l} 
\textbf{A}.\textbf{S} =  \widetilde{\textbf{A}}. \textbf{S} =0,\\ 
\textbf{A}.\widetilde{\textbf{S}}= 
\widetilde{\textbf{A}}. \widetilde{\textbf{S}} =0. 
\end{array} 
\right. 
\end{equation} 

When the wave is \textit{elliptically}-polarized, the 
amplitude bivector \textbf{A} is such that $\textbf{A} \times
\widetilde{\textbf{A}} \neq \mathbf{0}$ \cite{Haye84}. 
Then we deduce from \eqref{A.S} that \textbf{S} and 
$\widetilde{\textbf{S}}$ are parallel (because  they are both 
parallel to $\textbf{A} \times \widetilde{\textbf{A}}$). 
This is only possible when \textbf{S} has real direction 
\cite{BoHa93}, which means that $\textbf{S}= k \textbf{n}$, 
where $k$ is some complex scalar and \textbf{n} is a real vector 
in the common direction of \textbf{S} and 
$\widetilde{\textbf{S}}$. 
In this case, the plane wave described by \eqref{x} is 
\textit{homogeneous}. 

On the other hand, when the wave is \textit{inhomogeneous}, the 
slowness bivector \textbf{S} is such that $\textbf{S} \times
\widetilde{\textbf{S}} \neq \mathbf{0}$. 
Then by \eqref{A.S}, $\textbf{A}= \alpha \textbf{a}$, 
where $\alpha$ is a complex scalar and \textbf{a} is a real 
vector in the direction of polarization. 
In this case, the  wave is \textit{linearly}-polarized. 

Hence, if the wave is \textit{elliptically}-polarized  
($\textbf{A} \times \widetilde{\textbf{A}} \neq \mathbf{0}$) 
then it must be \textit{homogeneous};
 if the wave is \textit{inhomogeneous} 
($\textbf{S} \times \widetilde{\textbf{S}} \neq \mathbf{0}$) 
then it must be \textit{linearly}-polarized.

We conclude that, in an incompressible material, single trains 
of \textit{elliptically}-polarized \textit{finite}-amplitude 
\textit{inhomogeneous} plane waves of exponential type
may not propagate.

\medskip
\textit{Remark 1: Deformed material.}

Here, we assume the  incompressible material to have been first
subjected to a finite homogeneous static tri-axial stretch, with
stretch ratios $ \lambda_1, \lambda_2, \lambda_3 $ 
(and $ \lambda_1 \lambda_2 \lambda_3 = 1 $ to satisfy the 
incompressibility constraint). 
Upon this deformation, the inhomogeneous wave was then superposed.
Thus, a particle at \textbf{X} in the reference configuration has 
moved first to $\overline{\textbf{x}}$ given by
$\overline{x}_i= \lambda_i X_i  \, (i=1, 2, 3)$, and then to 
\textbf{x} given by
\begin{equation} \label{x2} 
\textbf{x} = \overline{\textbf{x}} 
+ \{ \textbf{A} e^{i\omega(\textbf{S.} \overline{\textbf{x}}-t)} 
+  \widetilde{\textbf{A}} 
e^{-i\omega(\tilde{\textbf{S}}.\overline{\textbf{x}}-t)} \}. 
\end{equation}

The deformation gradient associated with the deformation
\eqref{x2} is  
\begin{equation} \label{HF} 
\frac{\partial \textbf{x}} {\partial\textbf{X}} 
=\frac{\partial \textbf{x}}{\partial \overline{\textbf{x}}} \, 
 \frac{\partial \overline{\textbf{x}}}{\partial\textbf{X}}= 
\textbf{H} \,\mathrm{Diag} \, 
(\lambda_1, \lambda_2, \lambda_3), 
\end{equation} 
where the tensor $\textbf{H}$ is defined by 
\begin{equation} 
\textbf{H}= \textbf{1} + \omega \{ i \textbf{A} \otimes \textbf{S} 
e^{i \omega ( \textbf{S}.\overline{\textbf{x}}-t)} -
i \widetilde{\textbf{A}} \otimes \widetilde{\textbf{S}}
e^{-i\omega(\tilde{\textbf{S}}.\overline{\textbf{x}}-t)}\}.  
\end{equation}
A computation of the determinant $\overline{J}$ (say) of the
deformation tensor \eqref{HF}, yields
\begin{align} \label{Jbar}
\overline{J} 
= & \, \mathrm{det}(\textbf{H}) \, (\lambda_1 \lambda_2 \lambda_3)
=  \mathrm{det}(\textbf{H})  
\nonumber \\ 
= & \, 1 
+  \omega \{ i(\textbf{A}.\textbf{S}) 
 e^{i\omega(\textbf{S}.\overline{\textbf{x}}-t)} 
-i (\widetilde{\textbf{A}}.\widetilde{\textbf{S}}) 
e^{-i\omega(\tilde{\textbf{S}}.\overline{\textbf{x}}-t)}\}  
\nonumber \\ 
& \qquad  
- \omega^2 \{ (\textbf{A}. \widetilde{\textbf{S}}) 
(\widetilde{\textbf{A}}. \textbf{S}) - (\textbf{A}. \textbf{S}) 
(\widetilde{\textbf{A}}. \widetilde{\textbf{S}}) \} 
 e^{i\omega(\textbf{S}-\tilde{\textbf{S}}).\overline{\textbf{x}}}. 
\end{align} 

Again, because of the  
constraint of incompressibility, we must have $\overline{J}=1$ at 
all times and equations \eqref{A.S} are recovered.

\medskip
\textit{Remark 2: Small deformations superposed on large.}

Note that in the context of \textit{small}-amplitude waves, terms 
of second order in the magnitude of the wave's amplitude are 
negligible when compared to terms of first order. 
The incompressibility constraint then yields 
(see equations \eqref{J} and \eqref{Jbar})
$\textbf{A.S}=0$, and does not prevent the wave from being both
elliptically-polarized and inhomogeneous.


\end{document}